\documentclass[letterpaper, 10 pt, conference]{ieeeconf}  

\IEEEoverridecommandlockouts                              
\usepackage{amssymb}
\usepackage{amsmath}
\usepackage{amsfonts}
\usepackage{mathtools}
\usepackage{graphicx}
\usepackage{algorithm}
\usepackage[noend]{algorithmic}
\usepackage{color} 

\newtheorem{theorem}{\textbf{Theorem}}[section]
 
\newtheorem{problem}[theorem]{\textbf{Problem}}

\newtheorem{proposition}[theorem]{\textbf{Proposition}}

\newcommand{\norm}[1]{|\!|#1|\!|}

\newcommand{\nn}{{\mathcal{N}\negthickspace\mathcal{N}}\negthinspace}
\newcommand{\R}{\mathbb{R}}
\newcommand{\N}{\mathbb{N}}

\newcommand{\pr}{\mathrm{Pr}}
\newcommand{\Int}{\mathrm{Int}}
\newcommand{\abs}{\mathrm{abs}}
\newcommand{\ct}{\mathrm{ct}}

\title{\LARGE \bf
NNSynth: Neural Network Guided Abstraction-Based Controller Synthesis for Stochastic Systems
}

\author{Xiaowu Sun and Yasser Shoukry
\thanks{
This work was partially sponsored by the NSF awards \#CNS-2002405 and \#CNS-2013824.
}
\thanks{
Xiaowu Sun and Yasser Shoukry are with Department of Electrical Engineering and Computer Science, University of California, Irvine {\tt\small \{xiaowus,yshoukry\}@uci.edu}
}%
}

\begin{document}

\maketitle
\thispagestyle{empty}
\pagestyle{empty}

\begin{abstract}
In this paper, we introduce NNSynth, a new framework that uses machine learning techniques to guide the design of abstraction-based controllers with correctness guarantees. NNSynth utilizes neural networks (NNs) to guide the search over the space of controllers. The trained neural networks are ``projected'' and used for constructing a ``local'' abstraction of the system. An abstraction-based controller is then synthesized from such ``local'' abstractions. If a controller that satisfies the specifications is not found, then the best found controller is ``lifted'' to a neural network for additional training. Our experiments show that this neural network-guided synthesis leads to more than $50\times$ or even $100\times$ speedup in high dimensional systems compared to the state-of-the-art.
\end{abstract}

\section{Introduction}
\label{sec:introduction}
Abstraction-based control synthesis techniques have gained considerable attention in the past decade. These techniques provide tools for automated, correct-by-construction controller synthesis from complex specifications, typically given in the form of a Linear Temporal Logic (LTL) formulae~\cite{tabuada2009verification}. It is then unsurprising the vast amount of developed software tools that can handle a wide variety of nonlinear control systems including Pessoa~\cite{mazo2010pessoa}, CoSyMa~\cite{mouelhi2013cosyma}, SCOTS~\cite{rungger2016scots}, QUEST~\cite{jagtap2017quest}, FAUST~\cite{soudjani2015fau}, StocHy~\cite{cauchi2019stochy}, and AMYTISS~\cite{lavaei2020amytiss}. 
At the heart of all these tools is the need to obtain discrete abstraction of continuous-time dynamical systems using various quantization methods for state and input spaces. The resulting discrete abstraction is then traversed to search for a feedback controller that conforms to the required LTL specification. While performing the search for the feedback controller over the quantized system is motivated by the availability of tools from the computer science literature that can find such controllers, a significant drawback is the vast number of combinations of quantized states and inputs that needs to be considered. The problem is exacerbated in high-dimensional state and input spaces, leading to the so-called \emph{curse of dimensionality}. 

Motivated by the recent success of machine learning techniques in efficiently searching over the space of feedback controllers (e.g., imitation learning and reinforcement learning), we ask the following question: \emph{Can machine learning techniques be used to accelerate the process of synthesizing abstraction-based controllers from LTL specifications?} On the one hand, machine learning techniques enjoy favorable scalability properties and eliminate the dependency on state-space quantization. On the other hand, these learning-based feedback controllers (or policies) do not come with the guarantee that they conform to the LTL specifications. This motivates the need to closely integrate the scalability of learning-based techniques with the provable guarantees provided by the abstraction-based techniques.

Toward this end, we propose NNSynth, a new framework for synthesizing abstraction-based controllers from LTL specifications. Unique to NNSynth is the use of machine learning techniques to train a neural network (NN) based controller, which will guide the synthesis of the final abstraction-based controller. 
%
%
The advantages of the proposed NN guided abstraction-based controller synthesis is multi-fold. First, it utilizes the empirically proven advantages of machine learning algorithms to search the space of feedback controllers without relying on expensive quantizations of state and input spaces. Second, it limits the search over the quantized spaces only to local control actions within the neighborhood of the controller proposed by the NN training. That is, our approach uses NN training to guide the search over the quantized abstract system and eliminates the need to consider all combinations of quantized states and inputs. Third, the use of neural networks to guide the design of the abstraction-based controller opens the door to encode the human's preferences for how a dynamical system should act. Such human's preference is crucial for several real-world settings in which a human user or operator interacts with an autonomous dynamical system~\cite{RohanHRI}. Current research found that human preferences can be efficiently captured using expert demonstrations and preference-based learning which can be hard to be accurately capture in the form of a logical formulae or a reward function~\cite{palan2019learning}. These advantages are demonstrated using several key applications showing that NNSynth scales more favorably compared to the state-of-the-art techniques while achieving more than $50\times$ or even $100\times$ speedup in high dimensional systems.

\noindent \textbf{Related Work.}
The closest results to our work are those reported in~\cite{anderson2020neurosymbolic,verma2019imitation} which proposes a neurosymbolic framework to train control policies that can be represented as short programs in a symbolic language while ensuring the generated policies are safe. Similar to our approach, the work in~\cite{anderson2020neurosymbolic,verma2019imitation} trains a NN controller, project it to the space of symbolic controllers, analyze the symbolic controller and lift it back to the space of NN policies for further training. Differently, our approach focuses on designing a finite-state, abstraction-based controller instead of short programs in a symbolic language. This difference (short programs versus finite-state controller) manifests itself in all the framework steps, particularly the NN training, projection, and lifting. We confine our focus on synthesizing finite-state controllers due to the extensive literature on analyzing such controllers in tandem with the controlled physical systems~\cite{tabuada2009verification}.
%
Another line of related work is reported in~\cite{weiss2018extracting,carr2020verifiable} which studies the problem of extracting a finite-state controller from a recurrent neural network controller. We note that our framework uses the NN policy to guide the search for abstraction-based controllers and not as the final produced controller. 

\section{Problem Formulation}
\label{sec:formulation}

\subsection{Notation}
Let $|X|$ be the cardinality of a set $X$ and $\Int(X)$ be the interior of a set $X$. We denote the set of real numbers, positive real numbers, and natural numbers by $\R$, $\R^+$, $\N$, respectively. Let $\norm{x}$ be the Euclidean norm of a vector $x \in \R^n$ and $x^\top$ be the transpose of $x \in \R^n$. Let the inner product of two functions $h_1: X \rightarrow \R^m$ and $h_2: X \rightarrow \R^m$ be defined as $\langle h_1, h_2 \rangle = \int_X h_1(x)^\top h_2(x) dx$, which induces a norm $\norm{h_1} = \sqrt{\langle h_1, h_1 \rangle}$. We use $\nabla J$ to denote the Fréchet gradient of a functional $J$, and use the big $O$ notation for upper bounds.

\subsection{Dynamical Model}
We consider discrete-time nonlinear dynamical systems of the form:
\begin{equation}
    \label{eq:dyn}
    x^{(t+1)} = f(x^{(t)}, u^{(t)}) + g(x^{(t)}, u^{(t)}),
\end{equation}
where $x^{(t)} \in X \subset \R^n$ is the state and $u^{(t)} \in U \subset \R^m$ is the control input at time step $t \in \N$. The dynamical model consists of the priori known nominal model $f$ and the unknown model-error $g$ capturing unmodeled dynamics. Both functions $f$ and $g$ are assumed to be locally Lipschitz continuous. As a well-studied technique to learn unknown functions from data, we assume the model-error $g$ can be learned using Gaussian Process (GP) regression~\cite{GP}. We use $\mathcal{G}\mathcal{P}(\mu_g, \sigma^2_g)$ to denote a GP regression model with the posterior mean and variance functions be $\mu_g$ and $\sigma^2_g$, respectively\footnote{In the case of a multiple output function $g$, i.e., $m>1$, we model each output dimension with an independent GP. We keep the notations unchanged for simplicity.}. Given the dynamical system~\eqref{eq:dyn} with the model-error $g$ learned by $\mathcal{GP}(\mu_g, \sigma^2_g)$, let $\tau: X \times X \times U \rightarrow [0, 1]$ be the corresponding conditional stochastic kernel. Specifically, given the current state $x \in X$ and input $u \in U$, the distribution $\tau(\cdot|x,u)$ is given by the Gaussian distribution $\mathcal{N}(f(x,u)+\mu_g(x,u), \sigma^2_g(x,u))$. 

We treat the nonlinear system~\eqref{eq:dyn}, with the model-error $g$ learned by $\mathcal{GP}(\mu_g, \sigma^2_g)$, as a continuous Markov Decision Process (MDP) denoted by a tuple $\Sigma \triangleq (X, U, \tau)$. We denote by $T(A | x, u)$ the transition probability of reaching a subset $A\! \subset\! X$ in one step from state $x\! \in\! X$ with input $u\! \in\! U$:

{\small
\begin{equation}
    \label{eq:integrate_A}
    T(A | x, u) = \int_A \tau(x^\prime|x,u) dx^\prime.
\end{equation}}%
This integral can be easily computed since $\tau(\cdot | x, u)$ is a Gaussian distribution. 


\subsection{Abstraction-based Controller and Specification} 
We consider to control the continuous MDP $\Sigma$ (i.e., the nonlinear system~\eqref{eq:dyn} with the model-error learned by GP) using abstraction-based controllers. 
An abstraction-based controller considers to partition the continuous state space $X \subset \R^n$ into a finite set of abstract states $\widehat{X} = \{q_1, \ldots, q_N\}$, where each abstract state $q_i \in \widehat{X}$ is an infinity-norm ball in $\R^n$. The partitioning satisfies $X = \bigcup_{q \in \widehat{X}} q$ and $\Int(q_i) \cap \Int(q_j) = \emptyset$ if $i \neq j$. We denote by $\lambda \in \R^+$ the pre-specified grid size used for partitioning the state space. Let $\abs: X \rightarrow \widehat{X}$ map a state $x \in X$ to the abstract state $q = \abs(x)$ that contains $x$, i.e., $x \in q$, and $\ct: \widehat{X} \rightarrow X$ map an abstract state $q \in \widehat{X}$ to its center $\ct(q) \in X$, which is well-defined since abstract states are inifinity-norm balls. With some abuse of notation, we denote by $q$ both an abstract state, i.e., $q \in \widehat{X}$, and a subset of states, i.e., $q \subset X$.

Given a partitioning of the state space, an abstraction-based controller $\Psi: X \rightarrow U$ assigns the same control input to all states in the same abstract state, i.e., $\Psi(x_1) = \Psi(x_2)$ if $\abs(x_1) = \abs(x_2)$.
We denote by $\mathcal{S}$ the set of all abstraction-based controllers, where the underlying partitioning of the state space can be different for different abstraction-based controllers in $\mathcal{S}$. 


 For the high-level specifications, though our framework can be easily extended to general Linear Temporal Logic (LTL) specifications in bounded time horizon, we describe our algorithms with safety and liveness specifications for simplicity. Let $\xi_{x_0, \Psi}: \{1, \ldots, H\} \rightarrow X$ denote a closed-loop trajectory of~\eqref{eq:dyn} that starts from the state $x_0 \in X$ and evolves under the control law $\Psi$ for a bounded time horizon $H$. Then, we use $\xi_{x_0, \Psi} \models \phi_{\text{safety}}$ and $\xi_{x_0, \Psi} \models \phi_{\text{liveness}}$ to denote a trajectory $\xi_{x_0, \Psi}$ satisfying the safety and liveness specifications, respectively, i.e.,
 
\vspace{-3mm}
{\small
\begin{align*}    
    &\xi_{x_0,\Psi} \models \phi_{\text{safety}} \Longleftrightarrow \forall t \in \{1,\ldots H\},\; \xi_{x_0, \Psi}(t) \not\in X_\text{obst},  \\
    &\xi_{x_0,\Psi} \models \phi_{\text{liveness}} \Longleftrightarrow \exists t \in \{1,\ldots H\},\; \xi_{x_0, \Psi}(t) \in X_\text{goal},
\end{align*}}%
where $X_\text{goal} \subset X$ and $X_\text{obst} \subset X$ represent the goal and the obstacles, respectively. Given a specification $\phi = \phi_\text{safety} \land \phi_\text{liveness}$, we denote by $\pr\left(\Sigma_\Psi \models \phi\right)$ the average probability that the continuous MDP $\Sigma$ controlled by $\Psi$ satisfies the specification $\phi$ (averaged over initial states). 


\begin{figure*}[!ht]
    \centering
    \includegraphics[width = 0.95\textwidth]{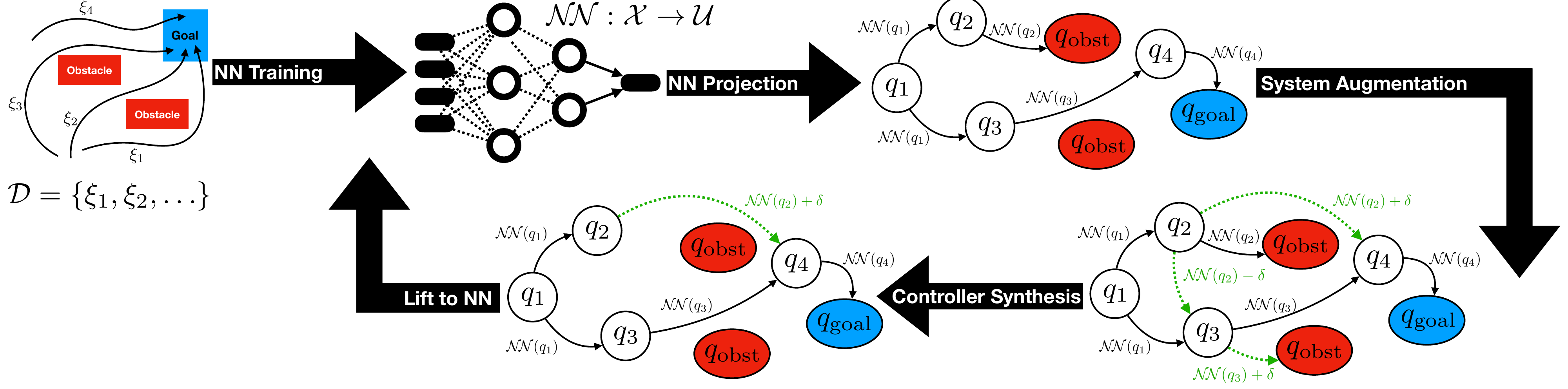}
    \caption{A cartoon summarizing the NNSynth framework. NNSynth starts by training a neural network controller $\nn$ using the data provided by the expert $\mathcal{D}$. The obtained neural network is then projected to an abstraction-based controller by evaluating the neural network on the representative points of abstract states, i.e. using the control actions $\nn(\ct(q))$. The obtained finite-state abstraction is then augmented with control actions in the neighborhood of the actions proposed by the neural network $\nn(\ct(x))\pm i \delta$. A controller is then synthesized from the augmented model. In case that a controller was not found, the ``best'' controller so far is then lifted to a neural network controller which is further trained using the expert data $\mathcal{D}$ to obtain a new $\nn$. The loop continues until an abstraction-based controller is found.} 
    \label{fig:arch}
\end{figure*}

\subsection{Main Problem}
The goal of this paper is to synthesize an abstraction-based controller $\Psi \in \mathcal{S}$ for the continuous system~\eqref{eq:dyn} to satisfy the given specifications $\phi$ while minimizing some given cost. The cost functional of a controller $\Psi$ is defined as $J(\Psi) = \int_X c(x, \Psi(x)) d\mu^\Psi(x)$, where $c(x, u)$ is the state-action cost and $\mu^\Psi$ is the distribution of states induced by the controller $\Psi$. Now, we can define the problem of interest as follows:
\begin{problem}
    \label{prob:main} 
    Given a cost functional $J$, a high-level specification $\phi$ and a user defined threshold $p$, we are interested in synthesizing an abstraction-based controller $\Psi: X \rightarrow U$ for the continuous MDP $\Sigma$ to minimize the cost $J(\Psi)$ while satisfying the specification $\phi$ with probability at least $p$:
    \begin{align}
        \Psi^* =\underset{\Psi \in \mathcal{S}}{\text{argmin}}\ J(\Psi) \quad
        \text{s.t.}\ \pr \left(\Sigma_\Psi \models \phi \right) \geq p. 
    \end{align} 
\end{problem}

\section{NNSynth Framework}
\label{sec:framework}

Our framework is featured by the use of neural networks to guide the search of abstraction-based controllers satisfying the specification $\phi$, and the ability to utilize policy gradient approaches for abstraction-based controllers when minimizing the cost functional $J$. In this section, we first give an overview of our framework, and then present each step separately in the following subsections. 

The overview of the proposed NNSynth is depicted in Figure~\ref{fig:arch}. 
Algorithm~\ref{alg:nnsynth} outlines the framework. 
After initializing an abstraction-based controller $\Psi_0$ (line~\ref{line:initialize_nn}-\ref{line:project1} in Algorithm~\ref{alg:nnsynth}), NNSynth lifts the abstraction-based controller $\Psi_k$ to a neural network $\nn_k$ through imitation learning of the data generated by $\Psi_k$ (line~\ref{line:lift} in Algorithm~\ref{alg:nnsynth}), updates the neural network through either imitation learning of the expert dataset $\mathcal{D}_\text{exp}$ or reinforcement learning (by providing the state-action cost function $c$ instead of expert data) with learning rate $\eta$ (line~\ref{line:update2} in Algorithm~\ref{alg:nnsynth}), and finally synthesizes a new abstraction-based controller $\Psi_{k+1}$ under the guidance of $\nn_{k+1}$ (line~\ref{line:project2} in Algorithm~\ref{alg:nnsynth}). This loop iterates until the satisfaction probability $V_\text{avg}$ is no less than the pre-specified threshold $p + \varsigma$ (line~\ref{line:compare_p} in Algorithm~\ref{alg:nnsynth}).

\begin{algorithm}[!t]
    \caption{\textsc{NNSynth} ($\mathcal{D}_\text{exp}$, $\phi$, $H$, $p$, $\varsigma$, $\eta$)}
    \label{alg:nnsynth}
    {\small
    \begin{algorithmic}[1]
        \STATE Initialize $\nn_\text{init}$ with random weights \label{line:initialize_nn}
        \STATE $\nn_\text{init} = \textsc{Update}(\nn_\text{init}, \mathcal{D}_\text{exp}, \eta)$ \label{line:update1}
        \STATE $\Psi_0, V_\text{avg} = \textsc{Project-by-Synth} (\nn_\text{init}, \phi, H)$ \label{line:project1}
        \FOR{$k = 0, \ldots, K-1$} \label{line:nnsynth_for_start}
            \IF{$V_\text{avg} \ge p + \varsigma$} \label{line:compare_p}
                \STATE \textbf{Return} $\Psi_k, V_\text{avg}$
            \ENDIF
            \STATE $\nn_k = \textsc{Lift}(\Psi_k)$ \label{line:lift}
            \STATE $\nn_{k+1} = \textsc{Update}(\nn_k, \mathcal{D}_\text{exp}, \eta)$ \label{line:update2}
            \STATE $\Psi_{k+1}, V_\text{avg} = \textsc{Project-by-Synth} (\nn_{k+1}, \phi, H)$ \label{line:project2}
        \ENDFOR
        \STATE \textbf{Return} $\Psi_K, V_\text{avg}$
    \end{algorithmic}  
    }
\end{algorithm}

\subsection{Step 1: NN Training}
Starting from the expert-provided trajectories $\mathcal{D} = \{\xi_1, \xi_2, \ldots\}$, we use imitation learning to train a neural network controller $\nn$ for the continuous MDP $\Sigma$. Alternatively, the NN controller can be trained by reinforcement learning, which requires the expert to provide the state-action cost $c: X \times U \rightarrow \R$ instead of the dataset $\mathcal{D}$. Neural networks are highly parameterized and can be updated using gradient based approaches $\nn_{k+1}\! =\! \nn_k - \eta \nabla J(\nn_k)$, where $\eta \in \R^+$ is the learning rate. The gradient $\nabla J(\nn_\theta)$ of a neural network parameterized by weights $\theta$ can be approximated using sampled trajectories:

{\small
\begin{equation}
    \label{eq:nn_gradient}
    \nabla J(\nn_\theta) \approx \frac{1}{M} \sum_{i=1}^M \sum_{t=1}^H \nabla_\theta \nn_\theta(u_{i, t} | x_{i, t}) \widehat{Q}_i^t
\end{equation}}%
where $M$ is the number of trajectories, $H$ is the bounded time horizon, and $\widehat{Q}_i^t$ is the estimated cost-to-go. We use the neural network to improve the controller's performance (i.e., minimizing the cost functional $J$) although the gradient of an abstraction-based controller, denoted by $\nabla J(\Psi)$, does not exist. Detailed optimality analysis is given in Section IV.

\subsection{Step 2: NN Projection}
\label{subsec:project}
Regardless of the use of imitation learning or reinforcement learning, the resulting neural network $\nn$ is not guaranteed to satisfy the specification $\phi$ and hence can not be used directly as a controller. Nevertheless, the neural network contains relevant control actions that can be used to obtain the final controller. To that end, NNSynth constructs a finite-state abstraction guided by $\nn$. Given a partitioning of the state space, we denote by $\widehat{X} = \{q_1, \ldots, q_N\}$ the corresponding set of abstract states, where the partitioning grid size $\lambda \in \R^+$ is determined based on the theoretical guarantees to be achieved (see Section~\ref{sec:guarantee}). Then, the finite-state abstraction induced by $\nn$ is given as a tuple $\widehat{\Sigma}^{\nn} \triangleq (\widehat{X}, \widehat{U}^\nn, \widehat{T}^\nn)$ with {\small $\widehat{X} = \{q_1, \ldots, q_N\}$, $\widehat{U}^\nn =  \{\nn(\ct(q)) \; | \; q \in \widehat{X}\}$}, and

{\small
\begin{align*}
    &\widehat{T}^\nn(q^\prime | q, u) = 
    \begin{cases}
    T( q^\prime | \ct(q), u) & \text{if } \; u = \nn(\ct(q)) \\
    0  & \text{otherwise},
    \end{cases} 
\end{align*}}%
where the transition probabilities $T( q^\prime | \ct(q), u)$ can be computed as~\eqref{eq:integrate_A}. 
In other words, the finite-state abstraction $\widehat{\Sigma}^{\nn}$ considers only one control action $\nn(\ct(q))$ at each abstract state $q$ and discards all other possible control actions. Computing such abstraction $\widehat{\Sigma}^{\nn}$ is straightforward and entails evaluating the NN controller at the center of each each abstract state and computing the transition probabilities associated with these actions.

\subsection{Step 3: System Augmentation}
\label{subsec:augmentation}
As shown in Figure~\ref{fig:arch}, the finite-state abstraction $\widehat{\Sigma}^{\nn}$ may contain transitions that violate the given specification $\phi$. This stems from the fact that $\widehat{\Sigma}^{\nn}$ considers only the actions taken by the trained network $\nn$. Therefore, the next step is to ``augment'' $\widehat{\Sigma}^{\nn}$ with additional transitions corresponding to control actions that are close to those taken by $\nn$. This augmentation will provide the controller synthesis algorithm with more freedom to choose other control actions. Given a precision $\delta \in \R^+$ and a range parameter $I \in \N$ ($\delta$ and $I$ are determined based on theoretical guarantees in Section~\ref{sec:guarantee}), we construct the augmented finite-state abstraction $\widehat{\Sigma}^{\nn+\delta} \triangleq (\widehat{X}, \widehat{U}^{\nn+\delta}, \widehat{T}_{\nn+\delta})$ with:

{\small
\begin{align}
    &\widehat{X} = \{q_1, \ldots, q_n\},\\
    &\widehat{U}^{\nn+\delta} =  \{\nn(\ct(q)) \pm i \delta \; | \; q \in \widehat{X},\; i = 0, 1, \ldots, I \}, \label{eq:u_hat} \\
    &\widehat{T}^{\nn+\delta}(q^\prime | q, u) = 
    \begin{cases}
        T( q^\prime | \ct(q), u) & \text{if } \; u \in \{ \nn(\ct(q)) \pm i \delta | \\ & \qquad \qquad i = 0,1, \ldots, I \} \\
        0  & \text{otherwise}, \label{eq:t_hat}
    \end{cases}
\end{align}}%
where with some abuse of notation, we use $\nn(\ct(q)) \pm i \delta$ to denote $\nn(\ct(q)) + [\pm i_1 \delta,  \pm i_2 \delta, \ldots, \pm i_m \delta]^\top$ with $i_1, i_2, \ldots i_m \in \{0, 1, \ldots, I\}$. In other words, the augmented abstraction $\widehat{\Sigma}^{\nn+\delta}$ takes into account all the control actions that are $\delta, 2 \delta, \ldots I \delta$ away from those given by the neural network $\nn$, where the distance is considered for each dimension of the control input $u \in \R^m$.

\subsection{Step 4: Controller Synthesis} 
\label{subsec:synth}
The next step is to synthesize a controller $\widehat{\Psi}$ for the augmented abstraction $\widehat{\Sigma}^{\nn+\delta}$ to satisfy the specification $\phi$. 
We emphasize that though the controller $\widehat{\Psi}: \widehat{X} \rightarrow U$ is synthesized for the finite-state abstraction $\widehat{\Sigma}^{\nn+\delta}$, it yields an abstraction-based controller $\Psi: X \rightarrow U$ for the continuous system $\Sigma$ as $\Psi(x) = \widehat{\Psi}(\abs(x))$, i.e., applying the same control action $\widehat{\Psi}(q)$ at all states $x \in q$, where $q \in \widehat{X}$. The difference in the probabilities of satisfying the specification $\phi$ for the finite-state abstraction $\widehat{\Sigma}^{\nn+\delta}$ controlled by $\widehat{\Psi}$ and the continuous MDP $\Sigma$ controlled by $\Psi$ can be bounded~\cite{lavaei2021survey} (see Section~\ref{sec:guarantee}). 

With the notations introduced above, let $\widehat{\Sigma}^{\nn+\delta}_{\widehat{\Psi}}$ be the finite-state abstraction $\widehat{\Sigma}^{\nn+\delta}$ controlled by $\widehat{\Psi}: \widehat{X} \rightarrow \widehat{U}^{\nn+\delta}$. Given the bounded time horizon $H$, we define the value function $V: \widehat{X} \times \{0, \ldots, H\} \rightarrow [0, 1]$ by letting $V(q, t)$ be the probability of satisfying the given specification $\phi$ in $H - t$ time steps when the system $\widehat{\Sigma}^{\nn+\delta}_{\widehat{\Psi}}$ starts from $q \in \widehat{X}$. Then, the average probability of satisfying the specification $\phi$ is given by:

{\small
\begin{equation}
    \label{eq:v_avg}
    V_\text{avg} \triangleq \pr \left(\widehat{\Sigma}^{\nn+\delta}_{\widehat{\Psi}} \models \phi \right) = \frac{1}{|\widehat{X}|} \sum_{q \in \widehat{X}} V(q, 0).
\end{equation}}%

Algorithm~\ref{alg:project} presents details on the abstraction-based controller synthesis, which summarizes Subsections~\ref{subsec:project},~\ref{subsec:augmentation}, and~\ref{subsec:synth}. To maximize the probability of satisfying $\phi_\text{liveness}$ (similarly for $\phi_\text{safety}$), we solve the following dynamic programming (DP) recursion:

{\small
\begin{align}
    Q_t(q, u) &= \sum_{q^\prime \in \widehat{X}} V_{t+1}^*(q^\prime)\; \widehat{T}^{\nn+\delta}(q^\prime | q, u)\label{eq:abst_Q} \\
    V_t^*(q) &= \max_{u \in \{ \nn(\ct(q)) \pm i \delta |  i = 0,  \ldots, I \}} Q_t(q, u) \label{eq:abst_V}
\end{align}}%
with the initial condition $V_H^*(q) = 1$ if $q \subseteq X_\text{goal}$ and $0$ otherwise, where $t = H-1, \ldots, 0$, and the transition probability matrix $\widehat{T}^{\nn+\delta}$ is given by~\eqref{eq:t_hat}. Critical to speedups of NNSynth is that entries $\widehat{T}^{\nn+\delta}(q^\prime | q, u)$ are nonzero only when {\small $u \in \{ \nn(\ct(q)) \pm i \delta | i = 0, \ldots, I \}$}, i.e., the control actions are close to that suggested by the neural network. This avoids computing all the transition probabilities $\widehat{T}(q^\prime | q, u)$, and searching for the optimal action in the whole discretized input space~\eqref{eq:abst_V}, which are the computational bottlenecks for abstraction-based controller synthesis. 

In Algorithm~\ref{alg:project}, NNSynth first computes entries of $\widehat{T}$ that are suggested by the neural network $\nn$ (line~\ref{line:compute_t_start}-\ref{line:compute_t_end} of Algorithm~\ref{alg:project}). In particular, line~\ref{line:u_near_nn} of Algorithm~\ref{alg:project} checks if control action $u$ is close to the action given by $\nn$, and computes the corresponding entries of $\widehat{T}$ only if $u$ has not been considered before at $q$, i.e., $u \not \in U_\text{buffer}(q)$. The optimal control action at each state is determined by maximizing the Q-function (line~\ref{line:max_q_start}-\ref{line:max_q_end} in Algorithm~\ref{alg:project}). Unique to NNSynth, it only searches the local action space that contains $\nn(q, t)$ at $q$ (line~\ref{line:local_search} in Algorithm~\ref{alg:project}). Since the optimal policy is in general time-dependent, we explicitly include the time steps $t$ in the input feature to the neural network. In line~\ref{line:cutoff1} and~\ref{line:cutoff2} of Algorithm~\ref{alg:project}, $\widehat{B}_\rho(f(\ct(q), u))$ denotes the subset of abstract states that are in a ball centered at $f(\ct(q), u)$ with radius $\rho$, where $\rho$ is a user-provided probability cut-off (when probability is smaller than the cut-off, the probability is treated as zero), which allows further speedup by limiting the transitions due to the model-error~\cite{lavaei2020amytiss}.

\begin{algorithm}[!t]
    \caption{\textsc{Project-by-Synth} ($\nn$, $\phi$, $H$)}
     \label{alg:project}
    {\small
    \begin{algorithmic}[1]
    \IF{$\phi == \text{Safety}$}
        \STATE $V(q, H) = 1$ for all $q \in \widehat{X}$
    \ELSE    
        \STATE $V(q, H) = 0$ for all $q \in \widehat{X}$
    \ENDIF
    \STATE $U_\text{buffer}(q) = \text{set}()$ for all $q \in \widehat{X}$
    \FOR{$t = H-1, \ldots, 0$}
    \FOR{$q \in \widehat{X} \setminus (\widehat{X}_\text{goal} \bigcup \widehat{X}_\text{obst})$} \label{line:compute_t_start}
            \FOR{$u \in \{\nn(\ct(q), t) \pm i \delta | i = 0, \ldots, I\} \setminus U_\text{buffer}(q)$} \label{line:u_near_nn}
                \STATE $U_\text{buffer}(q).add(u)$
                \STATE Compute $\widehat{T}(q^\prime | q, u)$ for all $q^\prime \in \widehat{B}_\rho(f(\ct(q), u))$ \label{line:cutoff1}
                \IF{$\phi \neq \text{Safety}$}
                    \STATE Compute transition prob. to the goal $\widehat{T}(X_\text{goal} | q, u)$
                    \STATE $\widehat{T}(q^\prime | q, u) = 0$ for all $q^\prime \in \widehat{X}_\text{goal}$
                \ENDIF
            \ENDFOR
        \ENDFOR \label{line:compute_t_end}
        
        \FOR{$q \in \widehat{X} \setminus (\widehat{X}_\text{goal} \bigcup \widehat{X}_\text{obst})$} \label{line:max_q_start}
            \STATE $V_{\max} = 0$
            \FOR{$u \in \{\nn(\ct(q), t) \pm i \delta | i = 0, \ldots, I\}$} \label{line:local_search}
                \STATE $Q(q, u) = \sum_{q^\prime \in \widehat{B}_\rho(f(\ct(q), u))}\widehat{T}(q^\prime | q, u) V(q^\prime, t+1)$ \label{line:cutoff2}
                \IF{$\phi \neq \text{Safety}$}
                    \STATE $Q(q, u) = Q(q, u) + \widehat{T}(X_\text{goal} | q, u)$
                \ENDIF
                \IF{$Q(q, u) > V_{\max}$}
                    \STATE $V_{\max} = Q(q, u)$ \\
                    \STATE $\widehat{\Psi}(q, t) = u$
                \ENDIF
            \ENDFOR
            \STATE $V(q, t) = V_{\max}$
        \ENDFOR \label{line:max_q_end}
    \ENDFOR    
       
        \STATE $V_\text{avg} = \frac{1}{|\widehat{X}|} \sum_{q \in \widehat{X}} V(q, 0)$
        
        \STATE $\Psi(x, t) = \widehat{\Psi}(\abs(x), t)$
        
        \STATE \textbf{Return} $\Psi, V_\text{avg}$
    \end{algorithmic}  
    } 
\end{algorithm} 

\subsection{Step 5: Lift to NN}
To further minimize the cost $J(\Psi_k)$, NNSynth ``lifts'' the abstraction-based controller $\Psi_k$ found in the previous step to a neural network $\nn_k$, which allows us to employ the well-developed deep policy gradient approaches to update the controller. Such lifting can be done by imitation learning with sampled trajectories of the continuous MDP $\Sigma$ controlled by $\Psi_k$. The obtained neural network is then used as an initialization for further training by either reinforcement learning or imitation learning of the expert dataset $\mathcal{D}$. In Section~\ref{sec:guarantee}, we analyze the performance of the synthesized controllers by taking into account the error due to the lift. This loop of training a NN, obtaining a local abstract model, synthesizing a controller, and lifting back to NN is then continued until a controller is found.

\section{Theoretical Analysis}
\label{sec:guarantee}

\subsection{Correctness Guarantees and Specification Satisfaction}
\label{subsec:sat_phi}
We provide theoretical guarantees of NNSynth in both satisfying the given specification $\phi$ and minimizing the cost functional $J$ in this section. The satisfaction of $\phi$ with pre-specified probability is correct-by-construction. In particular, the procedure \textsc{Project-by-Synth} (Algorithm~\ref{alg:project}) maximizes the probability for the finite-state abstraction $\widehat{\Sigma}^{\nn+\delta}_{\widehat{\Psi}}$ to satisfy $\phi$, and the difference in the satisfaction probability is bounded between the finite-state abstraction and the original continuous system~\cite[Theorem~2.1]{lavaei2021survey}:

{\small
\begin{align}
    \label{eq:prob_diff}
    \left|\pr\left(\widehat{\Sigma}^{\nn+\delta}_{\widehat{\Psi}} \models \phi\right) -  \pr\left(\Sigma_{\Psi} \models \phi\right) \right| \le \lambda H \mathcal{A} L_\tau,
\end{align}}%
where $\Psi(x) = \widehat{\Psi}(\abs(x))$, $\lambda$ is the grid size in partitioning the state space, $H$ is the bounded time horizon, $\mathcal{A}$ is the Lebesgue measure of the state space $X$, and $L_\tau$ is the Lipschitz constant of the stochastic kernel $\tau$. Therefore, we only need to set the margin $\varsigma = \lambda H \mathcal{A} L_\tau$ in Algorithm~\ref{alg:nnsynth} to ensure that the continuous MDP $\Sigma_\Psi$ satisfies $\phi$ with the pre-specified probability.
\begin{theorem}
    \label{thm:correct} 
    \hspace{-2.5mm}
    Consider Algorithm~\ref{alg:nnsynth} returns an abstraction-based controller $\Psi_k$ with average probability $V_\text{avg} > p+\varsigma$, where $\varsigma = \lambda H \mathcal{A} L_\tau$. Then, the continuous MDP $\Sigma$ controlled by $\Psi_k$ is guaranteed to satisfy the given specification $\phi$ with probability at least $p$, i.e., $\pr \left(\Sigma_{\Psi_k} \models \phi \right) \geq p$.
\end{theorem}
\begin{proof}
    This directly results from the definition of $V_\text{avg}$ in~\eqref{eq:v_avg} and the bound~\eqref{eq:prob_diff}.
\end{proof}

\subsection{Projection and Lift Error}
\label{subsec:error}
Now, we focus on the performance analysis of NNSynth, i.e., the optimality of controllers returned by Algorithm~\ref{alg:nnsynth} in terms of minimizing the cost functional $J$. To circumvent evaluating the gradient of abstraction-based controllers $\nabla J(\Psi_k)$, each iteration of Algorithm~\ref{alg:nnsynth} lifts an abstraction controller $\Psi_k$ to a neural network $\nn_k$ (line~\ref{line:lift} in Algorithm~\ref{alg:nnsynth}), updates the neural network using its gradient $\nabla J(\nn_k)$ (line~\ref{line:update2} in Algorithm~\ref{alg:nnsynth}), and projects the updated neural network $\nn_{k+1}$ back to the abstraction-based controller space $\mathcal{S}$ (line~\ref{line:project2} in Algorithm~\ref{alg:nnsynth}). In this subsection, we focus on the lift and projection procedures, and present the overall performance guarantee in the next subsection. 

Given a neural network $\nn_{k+1}$, the procedure \textsc{Project-by-Synth} (Algorithm~\ref{alg:project}) projects $\nn_{k+1}$ to an abstraction-based controller $\Psi_{k+1}$ while maximizing the probability of satisfying the specification $\phi$ at the meantime. This leads to the projection error compared the abstraction-based controller $\Psi_{k+1}^*$ that minimizes the distance to the neural network, i.e., $\Psi_{k+1}^* = {\text{argmin}}_{\Psi \in \mathcal{S}} \norm{\Psi - \nn_{k+1}}$. Let the Lipshitz constant of the neural network $\nn_{k+1}$ be $L_{nn}$, i.e., $\norm{\nn_{k+1}(x_1) - \nn_{k+1}(x_2)} \leq L_{nn} \norm{x_1 - x_2}$, $\forall x_1, x_2 \in X$, the following proposition upper bounds the difference between the abstraction-based controller returned by our projection procedure \textsc{Project-by-Synth} (Algorithm~\ref{alg:project}) and the actual minimizer of the distance to $\nn_{k+1}$.
\begin{proposition}
    \label{prop:projection_error}
     At an arbitrary iteration $k \in \{1, \ldots, K-1\}$ in Algorithm~\ref{alg:nnsynth}, let $\Psi_{k+1}$ be the abstraction-based controller returned by the procedure \textsc{Project-by-Synth} (line~\ref{line:project2} in Algorithm~\ref{alg:nnsynth}), and $\Psi_{k+1}^* = {\text{argmin}}_{\Psi \in \mathcal{S}} \norm{\Psi - \nn_{k+1}}^2$, where $\nn_{k+1}$ is the updated NN (line~\ref{line:update2} in Algorithm~\ref{alg:nnsynth}). Then, the difference between $\Psi_{k+1}$ and $\Psi_{k+1}^*$ is upper bounded as follows:
    \begin{equation}
        \label{eq:projection_error}
        \norm{\Psi_{k+1} - \Psi_{k+1}^*} = O \left(\delta I + \lambda L_{nn} \right).
    \end{equation}
\end{proposition}

In the above proposition, $\lambda$ is the grid size in partitioning the state space, $\delta$ and $I$ are the precision and range parameters in system augmentation, respectively (see~\eqref{eq:u_hat}). 

\begin{proof}
Since $\Psi_{k+1}, \Psi_{k+1}^* \in \mathcal{S}$, we evaluate their values at the centers $\ct(q) \in X$:
{\small
\begin{align}
    \label{eq:diff_fnc}
    &\norm{\Psi_{k+1} - \Psi_{k+1}^*} = \sqrt{\int_X \norm{\Psi_{k+1}(x) - \Psi_{k+1}^*(x)}^2 dx} \notag \\
    &= \sqrt{\sum_{q \in \widehat{X}} \mathcal{A}_q \norm{\Psi_{k+1}(\ct(q)) - \Psi_{k+1}^*(\ct(q))}^2} \notag \\
    &\leq \sqrt{\mathcal{A}} \max_{q \in \widehat{X}} \norm{\Psi_{k+1}(\ct(q)) - \Psi_{k+1}^*(\ct(q))}, 
\end{align}}%
where $\mathcal{A}_q$ and $\mathcal{A}$ are the Lebesgue measure of the abstract state $q$ and state space $X$, respectively. Consider an arbitrary abstract state $q \in \widehat{X}$. By Proposition~\ref{prop:upsilon} (in Appendix), since $\norm{\nn_{k+1}(x_1) - \nn_{k+1}(x_2)} \leq \lambda L_{nn}$, $\forall x_1, x_2 \in q$, there exists $y \in q$ such that:

{\small
\begin{equation}
    \label{eq:third_term}
   \norm{\nn_{k+1}(y) - \Psi_{k+1}^*(\ct(q))} \leq \lambda L_{nn}.
\end{equation}}%
With this choice of $y$, we have:
{\small
\begin{align}
    \label{eq:diff_x}
    &\norm{\Psi_{k+1}^*(\ct(q)) - \Psi_{k+1}(\ct(q))} \notag \\
    &\leq \norm{\Psi_{k+1}^*(\ct(q)) - \nn_{k+1}(y)} + \norm{\nn_{k+1}(y) - \nn_{k+1}(\ct(q))} \notag \\
    &\quad + \norm{\Psi_{k+1}(\ct(q)) - \nn_{k+1}(\ct(q))}  \notag \\
    &\leq \sqrt{m} \delta I + 2 \lambda L_{nn},
\end{align}}%
where the last step uses~\eqref{eq:third_term} and $\norm{\Psi_{k+1}(\ct(q)) - \nn_{k+1}(\ct(q))} \leq \sqrt{m} \delta I$. The later equation holds since the procedure \text{Project-by-Synth} first evaluates $\nn_{k+1}$ at the centers of abstract states, and then augments local actions within the radius $\delta I$ in each of the $m$ dimensions of $U \subset \R^m$. Since~\eqref{eq:diff_x} holds for an arbitrary $q \in \widehat{X}$, substituting~\eqref{eq:diff_x} into~\eqref{eq:diff_fnc} yields~\eqref{eq:projection_error}.
\end{proof}

The \textsc{Lift} procedure (line~\ref{line:lift} in Algorithm~\ref{alg:nnsynth}) trains a neural network $\nn_k$ whose output is close to that of the abstraction-based controller $\Psi_k$. In particular, we use the training approach~\cite{yun2019memorize} to memorize the outputs of $\Psi_k$ at the centers of abstract states (i.e., $\nn_k$ and $\Psi_k$ have exactly the same outputs at the centers of abstract states). The following proposition provides an upper bound for the lift error.
\begin{proposition}
    \label{prop:lift_error}
    Consider the neural network $\nn_k$ is given by lifting an abstraction-based controller $\Psi_k$, i.e., $\nn_k = \textsc{Lift} (\Psi_k)$ (line~\ref{line:lift} in Algorithm~\ref{alg:nnsynth}), and the Lipschitz constant of $\nn_k$ is $L_{nn}$, then $\norm{\nn_k - \Psi_k} = O \left( \lambda L_{nn} \right)$.
\end{proposition}

\begin{proof}
By evaluating $\Psi_k$ at the centers $\ct(q)$, we have:
{\small
\begin{align}
    &\norm{\nn_k - \Psi_k} = \sqrt{\int_X \norm{\nn_k(x) - \Psi_k(x)}^2 dx}  \\
    &= \sqrt{\sum_{q \in \widehat{X}} \int_q \norm{\nn_k(x) - \Psi_k(\ct(q))}^2 dx} \leq \sqrt{\mathcal{A}} \lambda L_{nn}
\end{align}}%
where $\mathcal{A}$ is the Lebesgue measure of the state space; the last step uses $\norm{\nn_k(x) - \Psi_k(\ct(q))} \leq \norm{\nn_k(x) - \nn_k(\ct(q))} + \norm{\nn_k(\ct(q)) - \Psi_k(\ct(q))} \leq \lambda L_{nn} + c$, $\forall x \in q$, where {\small $c \triangleq \max_{q \in \widehat{X}} \norm{\nn_k(\ct(q)) - \Psi_k(\ct(q))}$} can be zero by training NN to memorize the outputs of $\Psi_k$ at all the centers of abstract states.
\end{proof}

\subsection{Overall Regret}
\label{subsec:regret}
In Algorithm~\ref{alg:nnsynth}, the procedure \textsc{Update} (line~\ref{line:update2} in Algorithm~\ref{alg:nnsynth}) improves the neural network $\nn_k$ using its gradient, i.e., $\nn_{k+1}\! =\! \nn_k - \eta \nabla J(\nn_k)$, where $\eta$ is the learning rate and $\nabla J(\nn_k)$ can be evaluated as~\eqref{eq:nn_gradient}. This can be treated as an approximation of updating the abstraction-based controller $\Psi_k$ directly through $\Upsilon_{k+1} = \Psi_k - \eta \nabla J(\Psi_k)$, where $\Upsilon_{k+1}: X \rightarrow U$ is not necessarily an abstraction-based controller and need to be projected back to the abstraction-based controller space $\mathcal{S}$. We take into account this gradient approximation error, along with the projection and lift errors in the previous subsection, to provide the overall performance guarantee of NNSynth in terms of regret as follows:
\begin{theorem}
    \label{thm:regret}
    Consider the loop (line~\ref{line:nnsynth_for_start}-\ref{line:project2}) in Algorithm~\ref{alg:nnsynth} executes $K$ iterations and the abstraction-based controller obtained at the end of each iteration is $\Psi_k$, $k = 1, \ldots, K$. Let $\Psi^*$ be the optimal abstraction-based controller, i.e., $\Psi^* = \text{argmin}_{\Psi \in \mathcal{S}} J(\Psi)\; \text{s.t.}\; \pr \left(\Sigma_\Psi \models \phi \right) \geq p$. Then, the regret over $K$ iterations is upper bounded as follows:
    
    {\footnotesize
    \begin{equation}
    \label{eq:regret}
    \frac{1}{K} \sum_{k=1}^{K} J(\Psi_k) - J(\Psi^*) = O \left( \frac{1}{\eta K} + \frac{\delta I + \lambda L_{nn}}{\eta} + \lambda L_{nn} + \eta \right).
    \end{equation}}%
\end{theorem}

In the above theorem, by choosing the learning rate {\small $\eta = \sqrt{\frac{1}{K} + \delta I + \lambda L_{nn}}$}, \eqref{eq:regret} becomes:
{\footnotesize
\begin{equation}
    \frac{1}{K} \sum_{k=1}^{K} J(\Psi_k) - J(\Psi^*) = O \left( \lambda L_{nn} +  \sqrt{\frac{1}{K} + \delta I + \lambda L_{nn}} \right), \notag
\end{equation}}%
which shows that when the precision parameters $\lambda$ and $\delta$ approach zero, the regret can be arbitrarily small by increasing the number of iterations $K$. In general, the choice of parameters $\lambda$ and $\delta$ depends on the satisfaction probability and regret that need to be achieved, and these parameters can be determined based on Theorem~\ref{thm:correct} and Theorem~\ref{thm:regret}.

In the proof, we use $D: \mathcal{U} \times \mathcal{U} \rightarrow \R$ to denote the distance between two controllers for simplicity of notation, i.e., $D(\Upsilon_1, \Upsilon_2) \triangleq \frac{1}{2} \norm{\Upsilon_1 - \Upsilon_2}^2$, $\forall \Upsilon_1, \Upsilon_2 \in \mathcal{U}$. We will use the following identity: for all $\Upsilon_1, \Upsilon_2, \Upsilon_3 \in \mathcal{U}$ it holds that
{\small
\begin{align}
    \label{eq:bregman_identity}
    \langle \Upsilon_1 - \Upsilon_2, \Upsilon_1 - \Upsilon_3 \rangle = D(\Upsilon_1, \Upsilon_2) + D(\Upsilon_3, \Upsilon_1) - D(\Upsilon_3, \Upsilon_2).
\end{align}}%

\begin{proof} 
Let $\beta_k$ be the error due to the gradient approximation using neural networks, i.e., {\small $\nn_{k+1} = \Upsilon_{k+1} + \beta_k$}, where {\small $\nn_{k+1}\! =\! \nn_k - \eta \nabla J(\nn_k)$} and {\small $\Upsilon_{k+1} = \Psi_k - \eta \nabla J(\Psi_k)$}. Due to the convexity of $J$ over $\mathcal{S}$, we have that for all $\Psi \in \mathcal{S}$:

{\small
\begin{equation}
    \label{eq:convex_j}
    J(\Psi_k) - J(\Psi) \leq \langle \nabla J(\Psi_k), \Psi_k - \Psi \rangle.
\end{equation}}%
We now bound the RHS:
{\small
\begin{align}
    &\langle \nabla J(\Psi_k), \Psi_k - \Psi \rangle = \frac{1}{\eta} \langle \Psi_k - \Upsilon_{k+1}, \Psi_k - \Psi \rangle  \\
    &= \frac{1}{\eta} \langle \Psi_k - \nn_{k+1}, \Psi_k - \Psi \rangle + \frac{1}{\eta} \langle \beta_k, \Psi_k - \Psi \rangle  \\
    &= \frac{1}{\eta} (D(\Psi, \Psi_k) - D(\Psi, \nn_{k+1}) +  D(\Psi_k, \nn_{k+1})) \notag \\ 
    &\quad + \frac{1}{\eta} \langle \beta_k, \Psi_k - \Psi \rangle \label{eq:rhs_identity} \\
    &\leq \frac{1}{\eta} (D(\Psi, \Psi_k) - D(\Psi, \Psi_{k+1}^*) - D(\Psi_{k+1}^*, \nn_{k+1}) \notag \\
    &\quad +  D(\Psi_k, \nn_{k+1})) + \frac{1}{\eta} \langle \beta_k, \Psi_k - \Psi \rangle \label{eq:rhs_minimizer} \\
    &\leq \frac{1}{\eta} (D(\Psi, \Psi_k) - D(\Psi, \Psi_{k+1}) + \varepsilon_k + \gamma_k) + \frac{1}{\eta} \langle \beta_k, \Psi_k - \Psi \rangle, \label{eq:rhs_final}
\end{align}}%
where the projection error {\small $\varepsilon_k \triangleq D(\Psi, \Psi_{k+1}) - D(\Psi, \Psi_{k+1}^*)$} and  {\small $\gamma_k \triangleq D(\Psi_k, \nn_{k+1}) - D(\Psi_{k+1}^*, \nn_{k+1})$} is the relative improvement. In the above,~\eqref{eq:rhs_identity} uses the identity~\eqref{eq:bregman_identity} above;~\eqref{eq:rhs_minimizer} is due to the generalized Pythagorean theorem: if {\small $\Psi_{k+1}^* = {\text{argmin}}_{\Psi^\prime \in \mathcal{S}} D(\Psi^\prime, \nn_{k+1})$}, then it holds that {\small $D(\Psi, \nn_{k+1}) \geq D(\Psi, \Psi_{k+1}^*) + D(\Psi_{k+1}^*, \nn_{k+1})$} for all {\small $\Psi \in \mathcal{S}$}. 

The projection error $\varepsilon_k$ can be bounded as follows:
{\small
\begin{align}
    &\varepsilon_k \triangleq D(\Psi, \Psi_{k+1}) - D(\Psi, \Psi_{k+1}^*) \label{eq:error_start} \\
    &\leq \langle \Psi_{k+1} - \Psi_{k+1}^*, \Psi_{k+1} - \Psi \rangle \label{eq:error_identity} \\
    &\leq \norm{\Psi_{k+1} - \Psi_{k+1}^*} \norm{\Psi_{k+1} - \Psi} \label{eq:error_cauchy} \\
    &\leq d \norm{\Psi_{k+1} - \Psi_{k+1}^*}, \label{eq:error_final} 
\end{align}}%
where the diameter {\small $d \triangleq \text{sup}_{\Psi, \Psi^\prime \in \mathcal{S}} \norm{\Psi - \Psi^\prime}$}. In the above,~\eqref{eq:error_identity} uses the identity~\eqref{eq:bregman_identity} and the fact that the distance defined by $D$ is nonnegative;~\eqref{eq:error_cauchy} is due to Cauchy–Schwarz inequality.
    
The relative improvement $\gamma_k$ can be bounded as follows:
{\small
\begin{align}
    &\gamma_k \triangleq D(\Psi_k, \nn_{k+1}) - D(\Psi_{k+1}^*, \nn_{k+1}) \\ 
    &= \frac{1}{2} \norm{\Psi_k}^2 - \frac{1}{2} \norm{\Psi_{k+1}^*}^2 + \langle \nn_{k+1}, \Psi_{k+1}^* - \Psi_k \rangle \label{eq:improve_divergence} \\
    &\leq\! \langle  \nn_{k+1} - \Psi_k, \Psi_{k+1}^* - \Psi_{k} \rangle - \frac{1}{2} \norm{\Psi_k - \Psi_{k+1}^*}^2 \label{eq:improve_convexity} \\
    &\leq \langle \Upsilon_{k+1}\! -\! \Psi_k, \Psi_{k+1}^*\! -\! \Psi_{k} \rangle\! -\! \frac{1}{2} \norm{\Psi_k\! -\! \Psi_{k+1}^*}^2\! +\! \langle \beta_k, \Psi_{k+1}^*\! -\! \Psi_k \rangle \nonumber  \\
    &\leq -\eta \langle \nabla J(\Psi_k), \Psi_{k+1}^*\! -\! \Psi_{k} \rangle\! -\! \frac{1}{2} \norm{\Psi_k\! -\! \Psi_{k+1}^*}^2\! +\! \langle \beta_k, \Psi_{k+1}^*\! -\! \Psi_k \rangle \nonumber \\
    %
    &\leq \frac{1}{2} \eta^2 L_J^2 + d \norm{\beta_k}, \label{eq:improve_final}
\end{align}}%
where $L_J$ is the Lipschitz constant of $J$;~\eqref{eq:improve_convexity} is due to the strong convexity {\small $\frac{1}{2}\norm{\Psi_{k+1}^*}^2 \geq \frac{1}{2}\norm{\Psi_k}^2 + \langle \Psi_k, \Psi_{k+1}^* - \Psi_{k} \rangle + \frac{1}{2} \norm{\Psi_k\! -\! \Psi_{k+1}^*}^2$};~\eqref{eq:improve_final} is because {\small $az - bz^2 \leq \frac{a^2}{4b}$, $\forall z \in \R$} and uses Cauchy–Schwarz inequality.

The error $\beta_k$ can also be bounded:
{\small
\begin{align}
    &\norm{\beta_k} = \norm{\nn_{k+1} - \Upsilon_{k+1}} \\
    &\leq \norm{\nn_k - \Psi_k} + \eta \norm{\nabla J(\nn_k) - \nabla J(\Psi_k)} \\
    &\leq (1 + \eta c_j) \norm{\nn_k - \Psi_k} \label{eq:beta_final}
\end{align}}%
where $c_j$ is the Lipschitz constant of {\small $\nabla J$}.

Substitute~\eqref{eq:error_final},~\eqref{eq:improve_final},~\eqref{eq:beta_final} into~\eqref{eq:rhs_final} yields:
{\small
\begin{align}
    &\langle \nabla J(\Psi_k), \Psi_k - \Psi \rangle \notag \\
    &\leq \frac{1}{\eta} (D(\Psi, \Psi_k) - D(\Psi, \Psi_{k+1}) + d \norm{\Psi_{k+1} - \Psi_{k+1}^*} \notag \\
    &\quad + 2d (1 + \eta c_j) \norm{\nn_k - \Psi_k}) + \frac{1}{2} \eta L_J^2. \label{eq:rhs_combine}
\end{align}}%
With~\eqref{eq:rhs_combine}, the summation of~\eqref{eq:convex_j} over $K$ iterations yields:
{\small
\begin{align}
    &\frac{1}{K} \sum_{k=1}^{K} J(\Psi_k) - J(\Psi) \leq \frac{1}{\eta K} (D(\Psi, \Psi_1) - D(\Psi, \Psi_{K+1})) \notag \\
    &+ \frac{d}{\eta} \norm{\Psi_{k+1} - \Psi_{k+1}^*} + \frac{2d}{\eta} (1 + \eta c_j) \norm{\nn_k - \Psi_k} + \frac{\eta}{2 \alpha} L_J^2. \label{eq:regret_intermediate}
\end{align}}%
By following the similar process as~\eqref{eq:error_start}-\eqref{eq:error_final}, we have {\small $D(\Psi, \Psi_1) - D(\Psi, \Psi_{K+1}) \leq d^2$}. By Proposition~\ref{prop:projection_error}, the projection error {\small $\norm{\Psi_{k+1} - \Psi_{k+1}^*} = O \left(\delta I + \lambda L_{nn} \right)$}. By Proposition~\ref{prop:lift_error}, the lift error {\small $\norm{\nn_k - \Psi_k} = O \left(\lambda L_{nn} \right)$}. With these bounds,~\eqref{eq:regret_intermediate} leads to~\eqref{eq:regret}.
\end{proof}

\section{Results}
\label{sec:result}

We implemented NNSynth in Python and evaluated its performance on a Macbook Pro 15 with 32 GB RAM and Intel Core i9 2.4-GHz CPU. To compare with existing tools, we run all experiments on a single CPU core without using GPUs to accelerate neural network training. 

\subsection{Benchmarks and Performance}
\label{subsec:benchmarks}
\begin{table} [!b]
    \caption{Comparison between NNSynth and AMYTISS.} 
    \begin{center} 
    \resizebox{.99\columnwidth}{!}{
    \begin{tabular}{|c|c|c|c|c|c|c|c|}
    \hline
    \textbf{Benchmark} & \textbf{2-d Robot} & \textbf{5-d Room Temp.} &\textbf{5-d Traffic} \\ 
    \hline\hline
    Specification $\phi$  & Reach-avoid    & Safety   & Safety  \\ \cline{1-4}
    Specification horizon $H$  & 16   & 8  & 7 \\ \cline{1-4}
    Problem complexity $|\widehat{X} \times \widehat{U}|$  & 705600    & 3429216   & $1.25 \times 10^8$\\ \cline{1-4} 
    Satisfaction Probability $V_{\text{avg}}$  & 96\% & 95\%  & 80\% \\ \cline{1-4}
    NNSynth (time) [s] & 49.0  & 319.1  & 367.7 \\ \cline{1-4}
    AMYTISS (time) [s] & 108.4 & 34640.0  & 23100.0 \\ \cline{1-4}
    Speedup & \textbf{2 x} & \textbf{108 x} &  \textbf{62 x} \\ \cline{1-4}
\end{tabular}       
}
\end{center}
\label{tab:comparison} 
\end{table}

We start by evaluating NNSynth on three benchmarks with an increasing number of complexity. We compare NNSynth with the state-of-the-art tool in synthesizing controllers for stochastic systems, AMYTISS~\cite{lavaei2020amytiss}.                              
Table~\ref{tab:comparison} summarizes the comparison results. For each of the benchmarks, we list the specification $\phi$ used in this experiment along with its horizon $H$, the complexity of the problem measured by the number of abstract states times the number of discretized control actions $|\widehat{X} \times \widehat{U}|$, the average probability of satisfying the specification (averaged over the state space) $V_{\text{avg}}$, the execution time for each of the two tools, and the corresponding speedup. Indeed, the last row in Table~\ref{tab:comparison} empirically proves that using neural networks to guide the controller synthesis provides significant improvement to the overall execution time. Below, we provide more details about each of the benchmarks. 

\begin{figure}[!t]
    \centering
    \includegraphics[width = 0.49\columnwidth]{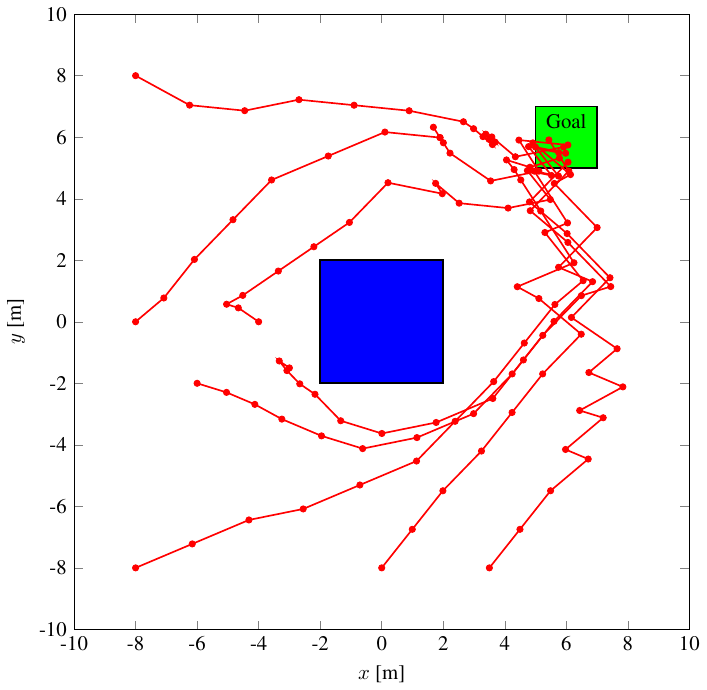} 
    \caption{Closed-loop trajectories sampled from different initial states using the synthesized controller in Experiment \#1.} 
    \label{fig:toy2d}
\end{figure}

\textbf{Experiment \#1: 2-d Robot.} 
Consider a 2-dimensional robot model given by:
{\small
\begin{align*}
    x_1^{(t+1)} &= x_1^{(t)} + u_1^{(t)} \text{cos}(u_2^{(t)}) + \varsigma_1^{(t)} \notag \\
    x_2^{(t+1)} &= x_2^{(t)} + u_2^{(t)} \text{sin}(u_2^{(t)}) + \varsigma_2^{(t)},
\end{align*}}%
where the state space $X = [-10, 10] \times [-10, 10]$, control input space $U = [-1, 1] \times [-1, 1]$, and the noise $(\varsigma_1, \varsigma_2)$ follows a Gaussian distribution with covariance matrix $\Sigma = \text{diag}(0.75, 0.75)$. We are interested in the task of steering the robot into a goal set $[5, 7] \times [5, 7]$ in $16$ time steps, while avoiding the obstacle set $[-2, 2] \times [-2, 2]$ (see Figure~\ref{fig:toy2d}). 

To construct the abstraction-based controller, we partition the state space with discretization parameters $(0.5, 0.5)$, and the input space with $(0.1, 0.1)$. This leads to a total number of $|\widehat{X}| = 1600$ abstract states and $|\widehat{U}| = 441$ control actions (by including the upper and lower limits of the input space as additional control actions) leading to a complexity of $|\widehat{X} \times \widehat{U}| = 705600$. NNSynth starts by training a neural network using imitation learning with a total of $121$ expert trajectories. The neural network consists of two hidden layers and ten neurons per hidden layer. We used Keras to train the neural network with the default adaptive learning rate optimization algorithm ADAM. By setting $\delta = 0.1$ (the same precision used to discretize the input space) and $I = 10$, NNSynth only needs to consider $100$ local control actions (out of the $|\widehat{U}| = 441$ total control actions) to construct the finite-state abstraction $\widehat{\Sigma}^{\mathcal{N\!N}+\delta}$. The controller synthesis is then executed to find a controller $\widehat{\Psi}$ that maximizes the probability of satisfying the specification, and one was found in $49.0$ seconds with an average satisfaction probability of $96\%$. The algorithm terminates in one iteration, and lifting the abstraction-based controller to a NN was not needed.

Using the same discretization parameters, AMYTISS was able to find a controller that satisfies the specs with $93\%$ probability in $108.4$ seconds. This shows a $2.2\times$ speedup of our tool (and an increase in the satisfaction probability) thanks to the fact that only $25\%$ of the state-action pairs are considered during the synthesis. These $25\%$ actions are chosen by the neural network that NNSynths used to guide the search. In Figure~\ref{fig:toy2d}, we present $8$ example trajectories under the control of $\widehat{\Psi}$, by sampling some initial states. 

\begin{figure*}[!t]  
  \center
  \resizebox{.95\textwidth}{!}{
    \includegraphics[]{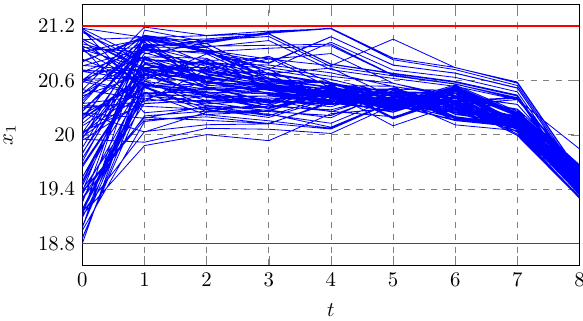}
    \includegraphics[]{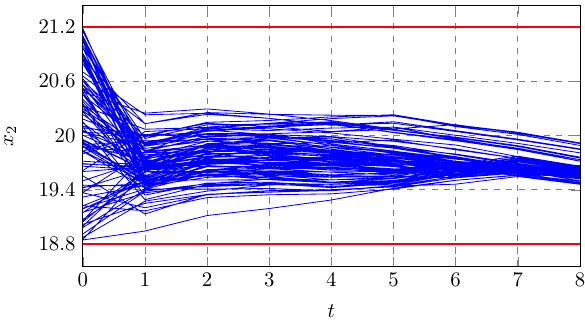}
    \includegraphics[]{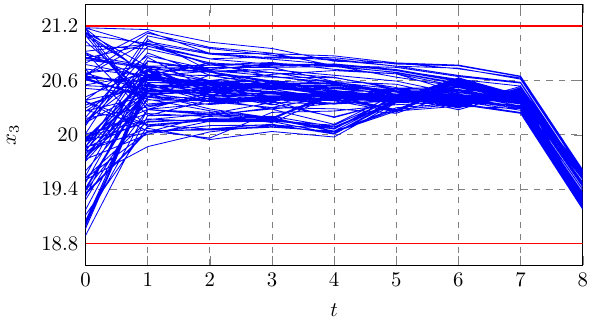}
    \includegraphics[]{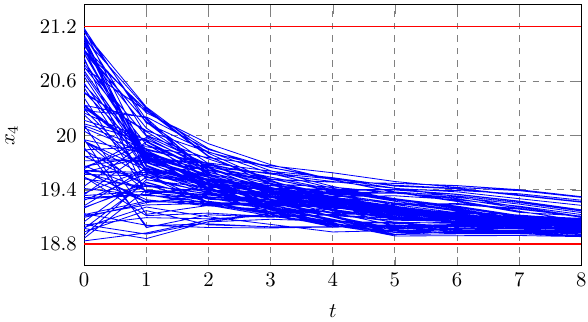}
    \includegraphics[]{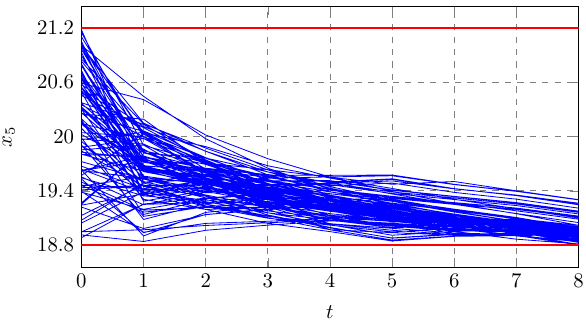}
  }
  \caption{State trajectories sampled from different initial conditions using the synthesized controller in Experiment \#2.}
  \label{fig:roomtemp5}  
\end{figure*}

\textbf{Experiment \#2: 5-d Room Temperature Control.} This example considers temperature regulation of 5 rooms each equipped with a heater and connected on a circle~\cite{lavaei2020amytiss}. The state variables are temperatures of individual rooms, and the evolution of the 5 room temperatures is described as:

{\small
\begin{align*}
    T_i^{(t+1)} &= a_{ii} T_i^{(t)} + \gamma T_h u_i^{(t)}\! +\! \eta w_i^{(t)} + \beta T_{ei}\! +\! 0.01 \varsigma_i^{(t)}, i \in \{1, 3\} \\
     T_i^{(t+1)} &= b_{ii} T_i^{(t)} + \eta w_i^{(t)} + \beta T_{ei} + 0.01 \varsigma_i^{(t)},\ i \in \{2, 4, 5\}
\end{align*}}%

\noindent where {\small $a_{ii} = (1-2\eta-\beta-\gamma u_i^{(t)})$}, {\small $b_{ii} = (1-2\eta-\beta)$}, and {\small $w_i^{(t)} = T_{i-1}^{(t)} + T_{i+1}^{(t)}$} (with {\small $T_0 = T_5$} and {\small $T_6 = T_1$}), and the parameters {\small $\eta=0.3$, $\beta = 0.022$, $\gamma = 0.05$, $T_{ei} = -1$, $T_h = 50$}.

We consider a safety specification that requires the temperature of each room to maintain in the safe set $[18.8, 21.2]$ for at least $8$ time steps. We partition the state space with grid size $0.4$ in each dimension, and use the grid size $(0.05, 0.05)$ for the input space $U = [0, 1] \times [0, 1]$. Similar to the previous benchmark, NNSynth trains a neural network with two layers and ten neurons per hidden layer using $935$ trajectories. With $I$ set to 7, NNSynth used only $49$ local control actions (out of $441$ total control actions) to compute the abstraction and synthesize a controller.
As shown in Table~\ref{tab:comparison}, NNSynth achieves a satisfaction probability of $95\%$ and $108\times$ speedup compared to AMYTISS. In Figure~\ref{fig:roomtemp5}, we sample $100$ initial states and present the evolution of the 5 state variables, which are all maintained within the safe set for at least $8$ steps under the abstraction-based controller provided by NNSynth.

\textbf{Experiment \#3: 5-d Road Traffic Network.}
This example considers a road traffic network divided into 5 cells, and state variables $x_i$ denote the number of vehicles per cell~\cite{lavaei2020amytiss}. The 5-d road traffic network is modeled as:
{\small
\begin{align*}
    x^{(t+1)}_1 &= (1-\frac{\tau v_1}{L_1}) x_1^{(t)} + \frac{\tau v_5}{L_5} w_1^{(t)} + 6 u^{(t)}_1 + 0.7\varsigma^{(t)}_1 \\
    x^{(t+1)}_i &= (1-\frac{\tau v_i}{L_i} - q) x_i^{(t)} + \frac{\tau v_{i-1}}{L_{i-1}} w_i^{(t)} + 0.7\varsigma^{(t)}_i,\ i \in \{2, 4\} \\
    x^{(t+1)}_3 &= (1-\frac{\tau v_3}{L_3}) x_3^{(t)} + \frac{\tau v_2}{L_2} w_3^{(t)} + 8 u^{(t)}_2 + 0.7\varsigma^{(t)}_3 \\
    x^{(t+1)}_5 &= (1-\frac{\tau v_5}{L_5}) x_5^{(t)} + \frac{\tau v_4}{L_4} w_5^{(t)} + 0.7\varsigma^{(t)}_5 
\end{align*}}%
where $w_i^{(t)} = x_{i-1}^{(t)}$ (with $x_0 = x_5$). Given the state space $X = [0, 10]^5$, the input space $U = [0, 1]^2$, a noise co-variance matrix $\Sigma = \text{diag}(0.7, 0.7, 0.7, 0.7, 0.7)$, and a probability cut-off 1e$-4$, we are interested in designing a control strategy that keeps the number of vehicles per cell in a safety set $[0, 10]$ for at least $7$ steps. To show the scalability of NNSynth, we partition the state space and the input space into $|\widehat{X}| = 12500$ and $|\widehat{U}| = 10000$ abstractions, respectively. This leads to a problem complexity in the order of $10^8$ control-action pairs. As shown in Table~\ref{tab:comparison}, NNSynth was able to solve this problem in $367.7$ seconds achieving more than $60 \times$ speedup compared with AMYTISS.

\subsection{Further Insights}
~\label{subsec:insights}
Beyond the performance evaluation, we conducted experiments to gain insights on the interaction between neural network training and abstraction-based controller synthesis. In particular, we aim to understand two questions: (i) how does the flexibility in the system augmentation (parameterized by $I$) help to discover the abstraction-based controller, and (ii) how does the abstraction-based controller help the neural network training?

\textbf{Experiment \#4: Effect of the parameter $I$ on performance}.
To answer the first question, we vary the number of local actions that are considered at each abstract state. To that end, Table~\ref{tab:local_actions} shows the result of running NNSynth with different values of $I$.
We report the probability of satisfying the specifications at the end along with the execution time.

In the 2-d robot case, the satisfaction probability grows from $55\%$ to $96\%$ by increasing the number of local state-control action pairs from $4$ to $100$. This shows that the neural network by itself is far away from the optimal control policy even if it is sufficiently trained. The reason behind this could be the neural network training is stuck at a local optimal. As favorable to NNSynth, abstraction-based controller synthesis can move away from the local optimal and further leads to better controllers, such as the one with satisfaction probability $96\%$ in the 2-d robot example. Similar pattern is observed in the other benchmark.

\textbf{Experiment \#5: Effect of lifting the abstraction controller to a neural network on performance}.
Now, consider the second question. 
We train the neural network for 50 epochs in each iteration and compare the satisfaction probabilities for the synthesized controllers after each iteration in Table~\ref{tab:iteration}. After five iterations of training, synthesizing a controller, lifting to a NN, and retraining, the NN receives a total of 250 epochs of training. For comparison,
we also record the base case where we train a neural network for only one iteration but with 250 epochs (the same total number of epochs as that accumulated over five iterations) but without lifting from the abstraction controller to the NN.

In the 2d Robot benchmark, the base case of 1 iteration with 250 epochs, the resulting controller achieved a $54\%$ probability of satisfying the specifications. On the other side, with several iterations of neural network training, controller synthesis, and lifting the controller to a neural network, the resulting controllers improve over iterations. By the end of the 5th iteration, the neural network accumulates 250 epochs (same as the base case), but the resulting satisfaction probability increases to $90\%$. This is a clear evidence that lifting the synthesized controller to a neural network helps with the overall training of neural networks. 

\begin{table}
    \caption{Numerical results for Experiment \#4.} 
    \begin{center} 
    \resizebox{.99\columnwidth}{!}{
    \begin{tabular}{|c|c|c|c|c|c|c|c|}
    \hline
    \textbf{Benchmark} & \textbf{Number of local control-} & \textbf{ Satisfaction} & \textbf{Execution time} \\ 
    \textbf{} & \textbf{action pairs ($I \times I$)} & \textbf{Probability $V_{\text{avg}}$} & \textbf{[s]} \\ 
    \hline\hline
               & 4  & 55\%  & 21.4 \\ \cline{2-4}
    2-d Robot  & 16   & 81\%   & 23.5\\ \cline{2-4} 
               & 49  & 90\%  & 31.2 \\ \cline{2-4}
               & 100  & 96\%  & 48.1 \\ \cline{1-4}
                    & 4   & 65\%  & 106.2  \\ \cline{2-4}
    5-d Room Temp.  & 16  & 94\% & 210.3 \\ \cline{2-4}
                    & 49  & 95\%  & 324.0 \\ \cline{2-4}
                    & 100 & 95\% & 630.9  \\ \cline{1-4}
\end{tabular}       
}
\end{center}
\label{tab:local_actions} 
\end{table}

\begin{table}
    \caption{Numerical results for Experiment \#5.} 
    \begin{center} 
    \resizebox{.99\columnwidth}{!}{
    \begin{tabular}{|c|c|c|c|c|c|c|c|}
    \hline
    \textbf{Benchmark} & \textbf{Iteration} & \textbf{Total training epochs} & \textbf{Satisfaction Probability $V_{\text{avg}}$}  \\ 
     & \textbf{number} & \textbf{at the end of iterations} & \textbf{at the end of iterations}  \\
    \hline\hline
                & (base case) 1  & 250   & 54\% \\ \cline{2-4} \cline{2-4}
                &  1  & 50   & 30\% \\ \cline{2-4}
                &  2  & 100  & 64\% \\ \cline{2-4}
    2-d Robot   & 3  & 150  & 82\% \\ \cline{2-4} 
                &  4  & 200  & 88\% \\ \cline{2-4}
                &  5  & 250  & 90\% \\ \cline{1-4} 
                
\end{tabular} 
}
\end{center}
\label{tab:iteration} 
\end{table}

\bibliographystyle{IEEEtran}
\bibliography{biblio.bib}


\appendix

\begin{proposition}
    \label{prop:upsilon}
     Let $\Upsilon: X \rightarrow U$ be an arbitrary controller and $\Psi_{\upsilon}^* = {\text{argmin}}_{\Psi \in \mathcal{S}} \norm{\Psi - \Upsilon}^2$. Consider an arbitrary abstract state $q \in \widehat{X}$. If $\exists c \in \R^+$ such that $\norm{\Upsilon(x_1) - \Upsilon(x_2)} \leq c$, $\forall x_1, x_2 \in q$, then $\exists y \in q$ such that $\norm{\Upsilon(y) - \Psi_\upsilon^*(\ct(q))} \leq c$.
\end{proposition}

\begin{proof}
By evaluating $\Psi$ at the centers $\ct(q)$, we have 
{\small
\begin{align}
    &\Psi_{\upsilon}^* = \underset{\Psi \in \mathcal{S}}{\text{argmin}} \norm{\Psi - \Upsilon}^2 = \underset{\Psi \in \mathcal{S}}{\text{argmin}} \int_X \norm{\Psi(x) - \Upsilon(x)}^2 dx \\
    &= \underset{\Psi \in \mathcal{S}}{\text{argmin}} \sum_{q \in \widehat{X}} \int_q \norm{\Psi(\ct(q)) - \Upsilon(x)}^2 dx.
\end{align}}%
Since the value of $\Psi_{\upsilon}^*(\ct(q))$ can be chosen independently in each $q$: 
{\small
\begin{equation}
    \label{eq:psi_upsilon_opt}
    \Psi_{\upsilon}^*(\ct(q)) = \underset{u \in U}{\text{argmin}} \int_q \norm{u - \Upsilon(x)}^2 dx.
\end{equation}}%
Now, we prove the proposition by contradiction. Assume that $\forall x \in q$, $\norm{\Upsilon(x) - \Psi_\upsilon^*(\ct(q))} > c$, then $\int_q \norm{\Upsilon(x) - \Psi_\upsilon^*(\ct(q))}^2 dx > c^2 \mathcal{A}_q$, where $\mathcal{A}_q$ is the Lebesgue measure of the abstract state $q$. This along with~\eqref{eq:psi_upsilon_opt} yields:

{\small
\begin{equation}
    \label{eq:min_greater_ca}
    \underset{u \in U}{\min} \int_q \norm{u - \Upsilon(x)}^2 dx > c^2 \mathcal{A}_q.
\end{equation}}%
Since $\norm{\Upsilon(x_1) - \Upsilon(x_2)} \leq c$, $\forall x_1, x_2 \in q$, by choosing $u = \Upsilon(x^\prime)$ with an arbitrary $x^\prime \in q$, we have $\int_q \norm{u - \Upsilon(x)}^2 dx \leq c^2 \mathcal{A}_q$, which contradicts~\eqref{eq:min_greater_ca}.
\end{proof}

\end{document}